\documentclass[a4paper,12pt]{article}
\usepackage{amsmath,amsthm,amssymb,bbm}

\usepackage[english]{babel}

\newcommand{\Ref}[1]{~(\ref{#1})}
\newcommand{\Cite}[1]{~\cite{#1}}

\def\greaterthansquiggle{\raise.3ex\hbox{$>$\kern-.75em\lower1ex\hbox{$\sim$}}}
\def\lessthansquiggle{\raise.3ex\hbox{$<$\kern-.75em\lower1ex\hbox{$\sim$}}}
\newcommand{\beq}{\begin{equation}}
\newcommand{\eeq}{\end{equation}}
\newcommand{\beqq}{\begin{eqnarray}}
\newcommand{\eeqq}{\end{eqnarray}}
\newcommand{\beqqn}{\begin{eqnarray*}}
\newcommand{\eeqqn}{\end{eqnarray*}}
\newcommand{\ba}{\begin{array}}
\newcommand{\ea}{\end{array}}

\newcommand{\ve}{\varepsilon}

\newcommand{\dg}{\dagger}
\newcommand{\wt}{\widetilde}
\newcommand{\wh}{\widehat}

\newcommand{\DI}{\mathbbm{D}}
\newcommand{\LI}{\mathbbm{L}}
\newcommand{\CI}{\mathbbm{C}}
\newcommand{\FI}{\mathbbm{F}}
\newcommand{\GI}{\mathbbm{G}}

\newcommand{\cS}{{\cal S}}

\def\nz{\ifmmode {I\hskip -3pt N} \else {\hbox {$I\hskip -3pt N$}}\fi}
\def\zz{\ifmmode {Z\hskip -4.8pt Z} \else
       {\hbox {$Z\hskip -4.8pt Z$}}\fi}
\def\qz{\ifmmode {Q\hskip -5.0pt\vrule height6.0pt depth 0pt
       \hskip 6pt} \else {\hbox
       {$Q\hskip -5.0pt\vrule height6.0pt depth 0pt\hskip 6pt$}}\fi}
\def\rz{\ifmmode {I\hskip -3pt R} \else {\hbox {$I\hskip -3pt R$}}\fi}
\def\cz{\ifmmode {C\hskip -4.8pt\vrule height5.8pt\hskip 6.3pt} \else
       {\hbox {$C\hskip -4.8pt\vrule height5.8pt\hskip 6.3pt$}}\fi}

\newtheorem{deft}{Definition}
\newtheorem{lem}{Lemma}

\newtheorem{prp}{Proposition}
\newtheorem{rem}{Remark}
\newtheorem{exa}{Example}
\def\au{{\setbox0=\hbox{\lower1.36775ex%
\hbox{''}\kern-.05em}\dp0=.36775ex\hskip0pt\box0}}
\def\ao{{}\kern-.10em\hbox{``}}

\begin{document}

\title{Asymptotic Entanglement and Lindblad Dynamics:\\ a Perturbative Approach}

\author{F. Benatti$^{a,b}$,
A. Nagy$^{c}$, H. Narnhofer$^{d}$\\
\small ${}^a$
Dipartimento di Fisica Teorica, Universit\`a di Trieste,
34014 Trieste, Italy\\
\small ${}^b$
Istituto Nazionale di Fisica Nucleare, Sezione di Trieste,
34014 Trieste, Italy\\
\small ${}^c$
Budapest University of Technology and Economics,
Muegyetem rkp.3-9, Hungary\\
\small ${}^d$
Fakult\"at f\"ur Physik, Universit\"at Wien}

\date{\null}

\maketitle
\begin{abstract}
We consider an open bipartite quantum system with dissipative dynamics generated by $\LI_\ve=\LI_0+\ve\LI_1$, where $\LI_{0,1}$ are generators of Lindblad type and $0<\ve<<1$.
In order to study the entanglement of the stationary states of $\LI_\ve$, we develop a perturbative approach and
apply it to the physically significant case when $\LI_0$ generates a reversible unitary dynamics, while $\LI_1$ is a purely dissipative perturbation.
\end{abstract}

\section{Introduction}
An open quantum system dynamics of Lindblad type\Cite{AL,BP} is an effective description of the action of an environment $E$ on a quantum system $S$ weakly coupled to it.
In general, the environment acts as a source of dissipation and noise; in spite of this, decoherence is not the only possible consequence. If suitably engineered, the coupling with the environment may also generate coherence and even entanglement\Cite{Hororev}, this possibility depending on the trade-off between dissipative effects and environment induced mixing\Cite{braun,beige,jakob1,jakob2,BFP,BLN,Isar}.

Of particular interest is under which conditions the presence of an environment may induce convergence to asymptotic states with definite entanglement properties\Cite{BF1,BF2,B,BLPal}. In fact, controlling the coupling to the environment could then be used for preparing states with definite entanglement content\Cite{Kraus}.
From this viewpoint, it is of great importance to know $1)$ the invariant states of a given Lindblad dynamics and $2)$ whether any initial state converges asymptotically to some stationary state.
A part from some older\Cite{Fri1,Fri2,spohn} and more recent results\Cite{FagnRebol,Dietz,HBW,HB}, a full characterization of the asymptotic properties of open quantum systems and their asymptotic behavior is still to be achieved.

In the following, we will focus upon the following scenario:
consider two finite-level systems $S_1$ and $S_2$, not directly interacting with each other, whose reversible, unitary dynamics
is generated by a Hamiltonian $H=H_1+H_2$ via the generator $\LI_0[\rho]=-i[H,\rho]$. If weakly coupled to a same environment, on a long time-scale, they undergo an open, dissipative dynamics generated by $\LI_\ve=\LI_0+\ve\,\LI_1$, where $\LI_1$ is a generator of Lindblad form and $\ve$ measures the weakness of the coupling to the environment.
In general, the addition of the perturbation term $\ve\,\LI_1$ diminishes the number of invariant states with respect to those of $\LI_0$; however, in finite dimension, at least one invariant state will always survive and, by continuity, will be close to them: the issue is whether the remaining invariant states may be entangled or not.

Suppose the spectrum of the Hamiltonian $H=H_1+H_2$ be non-degenerate, then $\LI_0[\rho_0]=0$ only if $\rho$ is a separable state. Intuitively, if such states are well inside the closed convex subset of separable states,
no dissipative perturbation $\LI_1$ could provide entangled states $\rho_\ve$ such that $\LI_\ve[\rho_\ve]=0$.
Indeed, by continuity, such asymptotic states are perturbations of those of $\LI_0$, namely $\rho_\ve=\rho_0+\ve\,\rho_1+o(\ve)$, and thus remain separable if $\rho_0$ is separable.
On the contrary, for separable stationary states $\rho_0$ on the boundary of the subset of separable states, it should be possible to construct entangled $\rho_\ve$ by suitably engineered, small dissipative perturbations.

In the following, we give mathematical ground to these expectations by developing
a systematic perturbation expansion of the states $\rho_\ve$ that are invariant under generators of the form  $\LI_\ve=\LI_0+\ve\LI_1$ where $\LI_0$ and $\LI_1$ are generic Lindblad type generators.

\section{Perturbation theory}

Let $S$ denote a $d$-level system with observables $X$ from the full matrix algebra $M_d(\mathbb{C})$ and states (density matrices) $\rho$ from the convex subset $\cS(S)\subset M_d(\CI)$ of positive matrices of trace $1$.
If $S$ is weakly coupled to its environment $E$, its time-evolution is conveniently approximated by a Markovian Lindblad-type dynamics: $\rho\mapsto\rho_t=\gamma_t[\rho]=\exp{(t\LI)}[\rho]$~\Cite{AL,BP} where $\LI$ is the generator of the semigroup of trace-preserving, completely positive maps $\gamma_t$, $\gamma_t\circ\gamma_s=\gamma_{t+s}$.
It incorporates in an effective manner the noise and dissipation due to the environment
via a master equation of the form
\beq
\label{genlind}
\partial_t\rho(t) =\LI[\rho(t)]=-i[H\,,\, \rho(t) ]\,+\,\DI[\rho(t)]\ ,
\eeq
where $M_d(\CI)\ni H=H^\dag$, while
\beq
\label{Kraus}
\DI[\rho]=\sum _{\alpha }\Big(
h_{\alpha }\,\rho\, h_\alpha^\dg\,-\,\frac{1}{2}\Big\{h_\alpha^\dg\,h_{\alpha }\,,\,\rho\Big\}\Big)\ ,
\eeq
where $h_\alpha\,,\,\sum_\alpha\,h^\dg_\alpha\,h_\alpha\in M_d(\CI)$.

We shall denote by $\cS_\gamma\subset\cS$ the subset of stationary states of $\gamma_t$: they satisfy $\LI[\rho]=0$ and form a convex subset of $\mathcal{K}(\LI)$, the kernel of the generator.
\medskip

The time-evolution generated by~(\ref{genlind}) affects the states of the system, while its observables evolve according to the semigroup of dual maps $\gamma^{\rm T}_t: X\mapsto X(t)
=\gamma^{\rm T}_t[X]$ generated by
\beq
\label{genlindual}
\partial_t X(t) =\LI^{\rm T}[X(t)]=i[H\,,\, X(t) ]\,+\,\DI^{\rm T}[\rho(t)]\ ,
\eeq
where
\beq
\label{Krausdual}
\DI^{\rm T}[X]=\sum _\alpha\Big(
h_\alpha^\dagger\,X\, h_\alpha\,-\,\frac{1}{2}\Big\{h_\alpha^\dg\,h_\alpha\,,\,X\Big\}\Big)\ .
\eeq

About the structure of the $\gamma_t$-invariant states, we have~\cite{Fri1}

\begin{prp}
\label{prop0}
The time-average
\beq
\label{timeav}
X\mapsto \GI^{\rm T}[X]=\lim_{T\to\infty}\frac{1}{T}\int_0^T{\rm d}t\,\gamma^{\rm T}_t[X]
\eeq
is a well-defined unital, completely positive map; its dual map $\GI:\cS\mapsto\cS_\gamma$ defined by
\beq
\label{dual}
{\rm Tr}\Big(\GI[\rho]\,X\Big)={\rm Tr}\Big(\rho\,\GI^{\rm T}[X]\Big)\qquad\forall X\in M_d(\CI)\ ,\ \rho\in\cS
\eeq
is a completely positive, trace preserving map which associates a state $\rho\in\cS$ to a $\gamma_t$-invariant state: $\LI\circ\GI[\rho]=0$.

If $\gamma_t$ possesses a faithful invariant state, that is a full-rank density matrix $\rho^*=\gamma_t[\rho^*]$,
then the $\gamma^{\rm T}_t$-invariant observables $X=\gamma^{\rm T}_t[X]$ form a subalgebra $M_\gamma\subset M_d(\CI)$ and $\GI^{\rm T}$ is a conditional expectation onto $M_\gamma$:
\beq
\label{cond-exp}
\GI^{\rm T}[Y_1\,X\,Y_2]=Y_1\,\GI^{\rm T}[X]\,Y_2\qquad\forall\ Y_{1,2}\in M_\gamma\ ,\ X\in M_d(\CI)\ .
\eeq
\end{prp}

Controlling the structure of $\cS_\gamma$ and, in particular, whether an initial state $\rho\in \cS$ converges to $\GI[\rho]$ is a complicated matter with some partial clues\Cite{Fri1,Fri2,spohn}. Recently, such an issue has become again subject of study\Cite{FagnRebol,Dietz,HBW,HB} also because of its increasing importance in quantum information\Cite{Kraus}.
Concerning $\gamma_t$-invariant states, the following result was obtained in~\cite{Umanita} (see also~\cite{HB}).
\medskip

\begin{prp}
\label{prop1}
Let $\LI$ be the generator of a Lindblad-type dynamics $\gamma_t$; one can always construct orthogonal stationary states $\rho_j$ of $\gamma_t$: $\LI[\rho_j]=0$ and $\rho_j\rho_k=0$ unless $j=k$.
\end{prp}
\medskip

The following ones are instances of some possible scenarios.
\medskip

\begin{exa}
\label{exa1}
Let $\LI[\rho]=-i[H,\rho]$, where the Hamiltonian $H=\sum_{j=1}^d E_j\,\vert j\rangle\langle j\vert$ has a non-degenerate spectrum; then, the stationary states $\rho_j$ of Proposition~\ref{prop1}
are the orthogonal one-dimensional eigen-projectors $\vert j\rangle\langle j\vert$.
For later application, we need extend the map $\GI$ of Proposition~\ref{prop0} to the whole matrix algebra $M_d(\CI)$; in the present case it reads
\beq
\label{ex3.3}
\GI[X]=\sum_{j=1}^d\langle j\vert\,X\,\vert j\rangle\,\vert j\rangle\langle j\vert\ .
\eeq
Clearly, because of the oscillatory behavior, there is no tendency to equilibrium: $\gamma_t[\rho]$ does not converge to $\GI[\rho]$ when $t\to+\infty$.
\end{exa}

\begin{exa}
\label{exa2}
Let $H=0$ in\Ref{genlind} and $h_\alpha=\vert\psi\rangle\langle \alpha\vert$ in\Ref{Kraus},
where $\{\vert\alpha\rangle\}$ is an orthonormal basis of $\CI^d$. Then,
\beq
\label{diss2}
\LI[\rho]=\vert\psi\rangle\langle\psi\vert\,-\,\rho\Longrightarrow
\gamma_t[\rho]={\rm e}^{-t}\,\rho+\Big(1-{\rm e}^{-t}\Big)\,\vert\psi\rangle\langle\psi\vert\ .
\eeq
Hence, $\rho=\vert\psi\rangle\langle\psi\vert$ is the only stationary state and all others converge to it asymptotically: $\gamma_t[\rho]\longmapsto\vert\psi\rangle\langle\psi\vert$.
The corresponding map $\GI$ is given by (on $M_d(\CI)$)
\beq
\label{ex3.2}
\GI[X]={\rm Tr}(X)\ \vert\psi\rangle\langle\psi\vert\ .
\eeq
\end{exa}

\begin{exa}
\label{exa3}
Let $H=0$ in\Ref{genlind} and $h_1=\vert\psi\rangle\langle 1\vert$ in\Ref{Kraus}, where $\|\psi\|=1$,
$\{\vert \alpha\rangle\}_{\alpha=1}^d$ is an orthonormal basis and
$0<|\psi_1|=|\langle1\vert\psi\rangle|<1$. Then,
\beq
\label{diss3}
\LI[\rho]=\langle1\vert\rho\vert1\rangle\,\vert\psi\rangle\langle\psi\vert
-\frac{1}{2}\vert1\rangle\langle1\vert\,\rho-\frac{1}{2}\rho\,\vert1\rangle\langle1\vert\ .
\eeq
Setting $X_{\alpha\beta}=\langle \alpha\vert X\vert\beta\rangle$ for all $X\in M_d(\CI)$,
$\mu=1-|\psi_1|^2>0$, $\psi_\alpha=\langle \alpha\vert\psi\rangle$, $\alpha\neq 1$, the equations,
$$
\dot{\rho}_{11}=-\mu\,\rho_{11}\ ,\quad
\dot{\rho}_{1\alpha}=\psi_1\psi^*_\alpha\,\rho_{11}\,-\,\frac{1}{2}\,\rho_{1\alpha}\ ,\quad
\dot{\rho}_{\alpha\beta}=\psi_\alpha\psi^*_\beta\,\rho_{11}
$$
can easily be solved yielding
\beqqn
&&
\rho_{11}(t)={\rm e}^{-\mu\,t}\,\rho_{11}\ ,\quad
\rho_{1\alpha}(t)={\rm e}^{-t/2}\,\rho_{1\alpha}\,+\,\psi_1\psi^*_\alpha\rho_{11}\frac{{\rm e}^{-t/2}-{\rm e}^{-\mu\,t}}{\mu-1/2}\\
&&
\rho_{\alpha\beta}(t)=\rho_{\alpha\beta}\,+\,\frac{1-{\rm e}^{-\mu\,t}}{\mu}\,\rho_{11}\,\psi_\alpha\psi^*_\beta\ ,
\eeqqn
where $a,b$ in the second expression are fixed by the initial conditions. Then,
\begin{equation}
\label{exa3.1}
\GI[X]=\sum_{\alpha,\beta\geq 2} \Big(X_{\alpha\beta}\,+\,\frac{X_{11}}{\mu}\psi_\alpha\psi^*_\beta\Big)\,\vert\alpha\rangle\langle\beta\vert\ .
\end{equation}
All states such that $\rho=Q\rho Q$, where $Q=\sum_{\alpha\geq 2}\vert\alpha\rangle\langle\alpha\vert$, are $\gamma_t$-invariant; also, $\gamma_t[\rho]\longmapsto\GI[\rho]$
\end{exa}

Despite the abstract characterizations of~\cite{HBW,HB}, the convex subset of stationary states is
difficult to control in practice; we shall thus concentrate on understanding how the invariant states
of a semigroup $\gamma^{(0)}_t$ are modified by a perturbation $\LI_1$ of its Lindblad generator $\LI_0$.
Concretely, we will investigate the set $\cS_\ve$ of stationary states of Lindblad-type dynamics $\gamma^{(\ve)}_t$ generated by $\LI_\ve=\LI_0+\ve\,\LI_1$, $0<\ve\ll1$.
By switching on the perturbation, the dimension of $\cS_\ve$ decreases, but, Proposition~\ref{prop0} ensures
the existence of at least one stationary state.

\begin{lem}
\label{lem1}
Consider the generator $\LI_\ve=\LI_0\,+\,\ve\,\LI_1$, where both $\LI_{0,1}$ are
generators in Lindblad form and the semigroups $\gamma_t^{(0)}$ and $\gamma_t^{(\ve)}$ generated by
$\LI_0$ and $\LI_\ve=\LI_0\,+\,\ve\,\LI_1$.
Let $n(0)$ be the number of $\gamma^{(0)}_t$-invariant orthogonal density matrices and $n(\ve)$
that of $\gamma^{(\ve)}_t$-invariant orthogonal density matrices for
$0<\ve\ll 1$; then $n(0)\,\geq\, n(\ve)$.
\end{lem}

\noindent
\textbf{Proof:}\quad
From Proposition~\ref{prop1}, one can always choose density matrices such that $\LI_\ve[\rho_j(\ve)]=0$ and
${\rm Tr}(\rho_j(\ve)\rho_k(\ve))=0$ for $j\neq k$.
In finite dimension, eigenvalues and eigen-projectors are continuous in $\ve$; therefore, should
$\rho_j(\ve)\neq \rho_k(\ve)$ merge as $\ve\to0$,
the continuity of the Hilbert-Schmidt scalar product would be violated.
\medskip

Because of finite dimensionality, the solutions can always be expressed as converging series in powers of $\ve$
\beq
\label{pertexp}
\rho_\ve=\sum_{n\geq 0}\ve^n\,\rho_n\ ,
\eeq
where the operators $\rho_n$ must solve the iterative procedure
\beqq
\label{pertexp0}
\Big(\LI_0+\ve\LI_1\Big)[\rho(\ve)]&=&\LI_0[\rho_0]+\sum_{n=1}\ve^n\Big(\LI_0[\rho_n]+
\LI_1[\rho_{n-1}]\Big)=0\ ,\quad \hbox{whence}\\
\label{pertexp1}
\LI_0[\rho_0]&=&0\ ,\qquad \LI_0[\rho_n]=-\LI_1[\rho_{n-1}]\quad n\geq 1\ ,
\eeqq
where $\rho_0$ is a stationary state of $\gamma^{(0)}_t$.
Also, since ${\rm Tr}(\rho(\ve))=1$, it follows that ${\rm Tr}(\rho_n)$ must vanish at all orders.
In the following, we discuss when $\rho_n=-\LI^{-1}_0[\LI_1[\rho_{n-1}]]$ are acceptable solutions.

\begin{deft}
\label{def0}
Let $\FI={\rm id}-\GI$, where $\GI$ is as in Proposition~\ref{prop0}; since $\GI$ is trace-preserving,
the image of $M_d(\CI)$ by $\FI$ consists of traceless matrices: ${\rm Tr}(\FI[X])=0$ for all $X\in M_d(\CI)$.
\end{deft}
\medskip

\begin{lem}
\label{lem2}
$\LI^{-1}$ can be defined as a map from $\FI[M_d(\CI)]$ into itself.
\end{lem}
\medskip

\noindent
\textbf{Proof:}\quad
Notice that $\GI$, as a time-average, maps into the kernel of $\LI$ and leaves it invariant; thus, from $X=\GI[X]+\FI[X]$, it follows that $\LI^{-1}[X]$ is well defined on $M_d(\CI)\ni X\neq 0$ only if $\GI[X]=0$.
Then, $\LI^{-1}$ is constructed as a linear map from the range of $\FI$ into itself such that $\LI\circ\LI^{-1}=\LI^{-1}\circ\LI={\rm id}$ on $\FI(M_d(\CI))$.
This guarantees that $\LI^{-1}[0]=0$; indeed, consider $Z=\LI^{-1}[X]-\LI^{-1}[Y]$, with
$X=\LI[V]=Y=\LI[V+W]$, $W\neq 0$, $\LI[W]=0$; then,
$\FI[Z]=\FI\circ\LI^{-1}[X]-\FI\circ\LI^{-1}[Y]=0$.
\medskip

\begin{exa}
\label{ex4}
In the case of Example~\ref{exa1}, where $\LI[\rho]=-i[H\,,\,\rho]$ and $H$ is non-degenerate,
$\GI[X]=0$ if and only if $\langle j\vert X\vert j\rangle=0$ for all $j$; one thus gets
\beq
\label{ex4.3}
\LI^{-1}[X]=i\sum_{j\neq k} \frac{\langle j\vert X\vert k\rangle}{E_j-E_k}\,\vert j\rangle\langle k\vert
\ .
\eeq
\end{exa}

\begin{exa}
\label{exa5}
In the case of Example~\ref{exa2}, $\GI[X]=0$ if and only if ${\rm Tr}(X)=0$;
one can verify that, on traceless matrices,
\beq
\label{ex5.1}
\LI^{-1}[X]=-X\ .
\eeq
\end{exa}

\begin{exa}
\label{exa6}
Finally, in Example~\ref{exa3}, $\GI[X]=0$ if and only if $X$ is of the form
$$
X=X_{11}\,\vert 1\rangle\langle 1\vert+\sum_{\alpha\geq 2}\Big(X_{1j}\,\vert 1\rangle\langle \alpha\vert+
X_{\alpha 1}\,\vert \alpha\rangle\langle 1\vert\Big)\ -
\frac{X_{11}}{\mu}\,Q\sum_{\alpha,\beta\geq 2}\psi_\alpha\psi^*_\beta\, \vert\alpha\rangle\langle\beta\vert\ ,
$$
where the only free entries are $X_{1\alpha}$ and $X_{\alpha1}$, $\alpha\geq 1$. Then,
\beqq
\nonumber
\LI^{-1}[X]&=&-\frac{1}{\mu}\Big(X_{11}\vert 1\rangle\langle 1\vert+\sum_{\alpha,\beta\geq2} X_{\alpha\beta}\,\vert\alpha\rangle\langle\beta\vert\Big)\\
\label{ex6.1}
&-&2\,
\sum_{\alpha\geq2}\Bigg(\Big(X_{1\alpha}-\frac{2}{\mu}\psi_1\psi^*_\alpha\Big)\,\vert1\rangle\langle\alpha\vert\,+\,
\Big(X_{\alpha1}-\frac{2}{\mu}\psi_\alpha\psi^*_1\Big)\,\vert\alpha\rangle\langle1\vert\Bigg)
\ .
\eeqq
\end{exa}
\medskip

From the previous Lemma, it follows that, in order to solve for $\rho_n$ in\Ref{pertexp1} by inverting $\LI_0$,
one has to ensure that $\GI_0[\LI_1[\rho_{n-1}]]=0$ for all $n\geq2$.
The following Lemma gives a sufficient condition for this to be true.
\medskip

\begin{lem}
\label{lem3}
Given $\LI_\ve=\LI_0+\ve\,\LI_1$, if $\LI_0[\rho_0]=0$ for a unique
state $\rho_0$ and $\LI_1[\rho_0]\neq 0$, then $\LI_\ve[\rho_\ve]=0$ for a unique $\rho_\ve$ given by
\beq
\label{aux0}
\rho_\ve=\sum_{n=0}^\infty(-\ve)^n\Big(\LI_0^{-1}\circ\LI_1\Big)^n[\rho_0]
=\frac{1}{1+\ve\,\LI_0^{-1}\circ\LI_1}[\rho_0]\ .
\eeq
\end{lem}

\noindent
\textbf{Proof:}\quad
As $\GI_0$ maps into the kernel of $\LI_0$ and is trace preserving, from the hypothesis of the lemma it follows that
$\GI_0\circ\LI_1[\rho_0]=\lambda\,\rho_0$. Then,
$\lambda={\rm Tr}\Big(\GI_0\circ\LI_1[\rho_0]\Big)={\rm Tr}\Big(\LI_1[\rho_0]\Big)=0$ implies
$\GI_0\circ\LI_1[\rho_0]=0$ so that $\LI_0$ can be inverted on $\LI_1[\rho_0]$ and one can solve the first recursive relation in\Ref{pertexp1}.
As $\LI^{-1}_0$ maps into $\FI_0[M_d(\CI)]$ where $\FI_0=1-\GI_0$ and $\GI_0$ is the trace-preserving
map in\Ref{timeav} corresponding to $\LI_0$, then ${\rm Tr}(\rho_1)=0$.
Iterating this argument yields the result.
\medskip

\begin{exa}
\label{ex5}
For $\LI_0$ as in Example~\ref{exa2}, there is only one invariant state so that Lemma~\ref{lem3} applies. Furthermore, since $\LI_0^{-1}[X]=-X$\Ref{aux0} yields
$\displaystyle\rho_\ve=\Big(1-\ve\LI_1\Big)^{-1}[\rho]$.
In such a case of a unique invariant state under $\LI_0$, we can make some preliminary considerations about the entanglement of the unique state invariant under $\LI_0+\ve\,\LI_1$.
Consider a separable pure state $\rho=P\otimes Q\in M_d(\CI)$, where $P=\vert\phi\rangle\langle\phi\vert$ and
$Q=\vert\chi\rangle\langle\chi\vert$; then, suitable non-local perturbations, $\LI_1$ may entangle it.
Indeed, by partial transposition\Cite{Hororev}, $\rho_\ve\mapsto\rho_\ve^\Gamma$, operated on the second party
with respect to an orthonormal basis starting with $\vert\chi\rangle$, one gets
\beq
\label{ent-gen0}
\rho_\ve^\Gamma=P\otimes Q+\ve\Big(\LI_1[P\otimes Q]\Big)^\Gamma\, +\,o(\ve)\ .
\eeq
By projecting  with $\Pi_\perp$ onto a subspace orthogonal to $P\otimes Q$, it follows that
$$
{\rm Tr}(\rho^\Gamma_\ve\,\Pi_\perp)=\ve\,{\rm Tr}\Big(\Pi_\perp\Big(\LI_1[P\otimes Q]\Big)^\Gamma\Big)\,
+\, o(\ve)\ .
$$
If $\LI_1[\rho]=-i[H_1\otimes 1+1\otimes H_2+H_{12},\rho]$, where $H_{12}$ is a non-local coupling of the two sub-systems, then the quantity
$$
{\rm Tr}(\rho_\ve\,\Pi_\perp)\simeq -i\ve\,{\rm Tr}\Big(\Pi_\perp\Big([H_{12}\,,\,P\otimes Q]\Big)^\Gamma\Big)
$$
can be made negative by suitably choosing $H_{12}$; then, one violates the positivity of partial transposition at order $\ve$ and $\rho_\ve$ is entangled at that order.

Entanglement can also be obtained via a purely dissipative time-evolution as the one generated by $\LI_1$ as in\Ref{diss3}; indeed, choosing $\vert 1\rangle\langle 1\vert=P\otimes Q$ yields
$$
{\rm Tr}(\rho^\Gamma_\ve\,\Pi_\perp)\simeq\ve\,{\rm Tr}\Big(\Pi_\perp\Big(\vert\psi\rangle\langle\psi\vert\Big)^\Gamma\Big)\ ,
$$
which can become negative by a suitable choice of entangled $\vert\psi\rangle$ and $\Pi_\perp$.
\end{exa}

The possibility of generating entanglement in the above two cases comes from the fact that the $0$-th order state $P_1\otimes P_2$ is on the border of the closed subset of separable states and can thus be moved into the  open complementary subset of entangled states by suitable terms of order $\ve$.

\subsection{dim(ker($\LI_0$))\,$\geq 2$}
\label{iterative-sec}

If, as in Examples~\ref{exa1} and~\ref{exa3}, the kernel of $\LI_0$ contains more than one stationary state, still
one may seek a $\rho_0$ such that $\LI_0[\rho_0]=0$ and
\beq
\label{pertexp2}
\GI_0\circ\LI_1[\rho_0]=\wh{\LI}_1[\rho_0]=0\ ,\qquad\hbox{where}\qquad
\wh{\LI}_1:=\GI_0\circ\LI_1\circ\GI_0\ ,
\eeq
so that the first order correction can be obtained as
$$
\rho_1=-\LI_0^{-1}\circ\LI_1[\rho_0]\ .
$$
In order to continue the iteration in~(\ref{pertexp0})
and get
$$
\rho_2=-\LI^{-1}_0\circ\LI_1[\rho_1]\ ,
$$
again by inverting $\LI_0$,
one has first to ensure that
\beq
\label{iter1}
\GI_0\circ\LI_1[\rho_1]=-\GI_0\circ\LI_1\circ\LI^{-1}_0\circ\LI_1[\rho_0]=0\ ,
\eeq
and, analogously, for the higher order contributions to\Ref{pertexp}.

\begin{exa}
\label{ex7}
Consider the case where $\LI_{0,1}[\rho]=-i[H_{0,1}\,,\,\rho]$ with $H_0$ non-degenerate.
With $H_0\vert E^0_j\rangle=E^0_j\vert E^0_j\rangle$ and using\Ref{ex3.3}, one gets
$$
\wh{\LI}_1[\rho]=-i\sum_{i,j=1}^d\langle E^0_i\vert\rho\vert E^0_i\rangle\
\langle E^0_j\vert\Big[H_1\,,\,\vert E^0_i\rangle\langle E^0_i\vert\Big]\vert E^0_j\rangle\
\vert E^0_j\rangle\langle E^0_j\vert=0
$$
for all $\rho$. Then, with
$\rho_0=\sum_{k=0}^dp_k\vert E^0_k\rangle\langle E^0_k\vert$, using\Ref{ex4.3}, one computes
$$
\rho_1=-\LI_0^{-1}\circ\LI_1[\rho_0]=-\sum_{j\neq k}\frac{p_k-p_j}{E^0_j-E^0_k}\,
\langle E^0_j\vert H_1\vert E^0_k\rangle\,\vert E^0_j\rangle\langle E^0_k\vert\ .
$$
From non-degenerate perturbation theory, the perturbation of
$\vert E^0_\ell \rangle$ to first order in $\ve$ is the eigenvector
$\vert\psi^{(\ell)}_\ve\rangle$ of $H_\ve=H_0+\ve\,H_1$ given by
$$
\vert\psi^{(\ell)}_\ve\rangle=\vert E^0_\ell\rangle+\ve\sum_{j\neq \ell}
\frac{\langle E^0_\ell\vert H_1\vert E^0_j\rangle}{E^0_\ell-E^0_j}\,\vert E^0_j\rangle\ .
$$
Thus one sees that, to order $\ve$, $\vert\psi^{(\ell)}_\ve\rangle\langle\psi^{(\ell)}_\ve\vert$
reproduces $\rho_1$ with $\rho_0=\vert E^0_\ell\rangle\langle E^0_\ell\vert$.
Furthermore,
$$
\GI_0\circ\LI_1[\rho_1]=i\sum_{j\neq k}\frac{p_j-p_k}{E^0_j-E^0_k}\,\left|\langle E^0_k\vert H_1\vert E^0_j\rangle\right|^2\,\Big(
\vert E^0_k\rangle\langle E^0_k\vert\,-\,\vert E^0_j\rangle\langle E^0_j\vert
\Big)=0\ .
$$
Thus, $\rho_2=\Big(\LI_0^{-1}\circ\LI_1\Big)^2[\rho_0]$ and so on with higher orders.
\end{exa}
\medskip

Unlike in the previous example, it may happen that\Ref{iter1} is not satisfied by the chosen $\rho_0$.
\medskip

\begin{exa}
\label{exa8}
Consider $\LI_0$ as in Example~\ref{exa1} and $\LI_1$ as in Example~\ref{exa2}: the solution to
$$
\LI_0[\rho]=0\quad \hbox{and to}\quad
\wh{\LI}_1[\rho]=\sum_{i,j=1}^d\rho_{jj}\,\langle i\vert\LI_1[\vert j\rangle\langle j\vert]\vert i\rangle\,\vert i\rangle\langle i\vert=\sum_{i=1}^d\Big(|\psi(i)|^2-\rho_{ii}\Big)\,\vert i\rangle\langle i\vert=0
$$
must have the form $\rho_0=\sum_{j=1}^d\,|\psi(j)|^2\,\vert j\rangle\langle j\vert$. Then, a natural candidate
for the first order perturbation contribution $\rho_1$ is, using $\LI^{-1}_0$ as in Example~\ref{ex4},
\beq
\label{exa8.1}
\rho_1=-\LI^{-1}_0\circ\LI_1[\rho_0]=-\LI^{-1}_0\Big[\vert\psi\rangle\langle\psi\vert-\rho_0\Big]=
-i\sum_{j\neq k}\frac{\psi(j)\psi^*(k)}{E_j-E_k}\, \vert j\rangle\langle k\vert\ .
\eeq
However, with this choice, it turns out that
$$
\GI_0\circ\LI_1[\rho_1]=\GI_0\Big[\vert\psi\rangle\langle\psi\vert-\rho_1\Big]=\sum_{j=1}^d|\psi(j)|^2\,\vert j\rangle\langle j\vert\neq 0\ .
$$
\end{exa}
\medskip

A possible strategy to overcome the problem exposed by the previous example is as follows:
in Lemma~\ref{lem2}, $\LI^{-1}_0$ is defined as a map from the range of $\FI_0$ into itself.
Thus, given a first order perturbation contribution
$\rho_1=-\LI_0^{-1}\circ\LI_1[\rho_0]$, one can always add to it (see the proof of the Lemma)
$\sigma_1\in M_d(\CI)$ such that $\LI_0[\sigma_1]=0$ whence
$\LI_0[\rho_1+\sigma_1]=\LI_0[\rho_1]=-\LI_1[\rho_0]$.
One can thus try to find an appropriate $\gamma^{(0)}_t$-invariant matrix $\sigma_1$ such that
$$
\GI_0\circ\LI_1[\rho_1]+\GI_0\circ\LI_1[\sigma_1]=\GI_0\circ\LI_1[\rho_1]+\wh{\LI}_1[\sigma_1]=0\ ,
$$
where we have used that $\GI_0[\sigma_1]=\sigma_1$. Thus, if such $\sigma_1$ can be found it is
of the form
\beq
\label{aitanos}
\sigma_1=-\wh{\LI}_1^{-1}\circ\GI_0\circ\LI_1[\rho_1]=\wh{\LI}^{-1}_1
\circ\GI_0\circ\LI_1\circ\LI_0^{-1}\circ\LI_1[\rho_0]\ ,
\eeq
where the inverse $\widehat{\LI}_1^{-1}$ of $\hat{\LI}_1$ is defined as in Lemma~\ref{lem2} and thus
${\rm Tr}(\sigma_1)=0$.
Then, one would obtain the second order perturbation contribution $\rho_2=-\LI_0^{-1}\circ\LI_1[\rho_1+\sigma_1]$.
\medskip

\noindent
\begin{rem}
\label{rem1}
Of course, the existence of $\sigma_1$ is equivalent to the invertibility of $\wh{\LI}_1$.
In general, $\wh{\LI}_1$ is not a generator of Lindblad form; namely, it does not generate a semigroup of completely positive maps on the set of all matrices, even if $\LI_1$ does.
However, in the next section, we shall consider the setting of Example~\ref{exa8} and
prove that $\wh{\LI}_1$ generate a positivity preserving semigroup on the density matrices commuting with $H$.
\end{rem}

\section{Dissipative Perturbation of a Unitary Evolution}

In the following, we restrict to a less general situation than the ones addressed in the previous section; namely,
we will stick to purely dissipative perturbations $\LI_1=\DI$ as in\Ref{Kraus} of
a generator $\LI_0[\rho]=-i[H\,,\rho]$ as in Example~\ref{exa1}:
\beq
\label{pertLind}
\LI_\ve[\rho]=-i[H\,,\,\rho]\,+\,\ve\,\DI[\rho]\ ,\quad
\DI[\rho]=\sum_\alpha\Big(
h_\alpha\,\rho\,h^\dag_\alpha-\frac{1}{2}\Big\{h^\dag_\alpha h_\alpha\,,\,\rho\Big\}\Big)\ .
\eeq

\begin{lem}
\label{lemm}
The map $\wh{\DI}=\GI_0\circ\DI\circ\GI_0$, with $\GI_0$ given by\Ref{ex3.3},
generates a positive, trace preserving map on the $\gamma^{(0)}_t$-invariant states $\rho$
that commute with $H$.
\end{lem}
\medskip

\noindent
\textbf{Proof:} If $\cS_0\ni\rho=\sum_{j=1}^d\rho_{jj}\,\vert j\rangle\langle j\vert$,
\beq
\label{condexp2}
\wh{\DI}[\rho]=\sum_{i,j=1}^d\rho_{ii}\,\langle j\vert\DI[\vert i\rangle\langle i\vert]\vert j\rangle\,\vert j\rangle\langle j\vert
=\sum_{i,j=1}^d\rho_{ii}\,\sum_\alpha\Big(|\langle j\vert h_\alpha\vert i\rangle|^2-\delta_{ij}\,\langle j\vert h^\dagger_\alpha h_\alpha\vert j\rangle\Big)\, \vert j\rangle\langle j\vert\ .
\eeq
Then,
\beq
\label{condexp2a}
\dot{\rho}_{jj}=\sum_{i=1}^d\rho_{ii}\,\sum_\alpha\Big(\Big|\langle j\vert h_\alpha\vert i\rangle\Big|^2-\delta_{ij}\,\langle j\vert h^\dagger_\alpha h_\alpha\vert j\rangle\Big)\geq
-h\,\rho_{jj}\ ,
\eeq
where $h=\Big\|\sum_\alpha h^\dag_\alpha h_\alpha\Big\|^2$.
Therefore, the eigenvalues of any $\rho\in\cS_0$ remain positive while evolving with $\exp(t\wh{\DI})$.
\medskip

\begin{exa}
\label{exa9}
Consider $\wh{\LI}_1$ as in Example~\ref{exa8}; on density matrices
$\cS_0\ni\rho=\sum_{j=1}^d\rho_{jj}\,\vert j\rangle\langle j\vert$ that commute with $H$, one finds
that $\exp(t\wh{\DI})[\rho]$ converges to the unique invariant state
$\rho_0=\sum_{j=1}^d|\psi(j)|^2\,\vert j\rangle\langle j\vert$. Indeed,\Ref{condexp2a} yields
$$
\dot{\rho}_{jj}=|\psi(j)|^2-\rho_{jj}\ ,\qquad \rho_{jj}(t)=|\psi(j)|^2\Big(1-{\rm e}^{-t}\Big)
+\rho_{jj}{\rm e}^{-t}\ .
$$
\end{exa}
\medskip

As already observed, even if one knows the structure of the invariant states of $\LI(\ve)$, it remains to be
proved that one actually has asymptotic convergence to them.
As showed in Lemma~\ref{lem1}, even a very weak perturbation $\ve\DI$ of $\LI_0$ in general decreases
the dimension of the kernel of $\LI_0$; this is why adding a suitable engineered dissipative perturbation can be used
to drive a system into a certain stationary state which may even be chosen to be pure\Cite{Kraus}.
However, one must check that all other eigenvalues of $\LI_\ve$ get a negative real part.
This cannot hold in general~\cite{HB}: purely imaginary eigenvalues can remain, but only if the Lindblad dynamics becomes trivial when reduced to the subspace which supports the stationary states. The following Lemma provides sufficient conditions for this not to be the case.
\medskip

\begin{lem}
\label{noimaginary}
Assume that, given the generator~(\ref{pertLind}), no projection $P\neq \mathbbm{1}$ can satisfy
$1)$ $[P,H]=[P,h_\alpha]=0$ and $2)$ $P\,h_\alpha\,P =c_\alpha\,P$ for all $\alpha$.
Then, the non-zero eigenvalues of $\LI_\ve$ have a strictly negative real part.
\end{lem}
\smallskip

\noindent
\textbf{Proof:}\quad
The matrix units $\vert j\rangle\langle k\vert$, $j\neq k$, are such that
$$
\LI_0[\vert j\rangle\langle k\vert]=-i(E_j-E_k)\,\vert j\rangle\langle k\vert\ .
$$
Consider $\LI_\ve[\rho_{jk}(\ve)]=\lambda_{jk}(\ve)\,\rho_{jk}(\ve)$ and the
expansion of eigen-matrices $\rho_{jk}(\ve)=\rho^{(0)}_{jk}+\ve\rho^{(1)}_{jk}$ and eigenvalues
$\lambda_{jk}(\ve)=-i(E_j-E_k)+\ve\,\eta_{jk}$ to first order in $\ve$. Inserted into the eigenvalue equation, this yields
$$
\LI_0[\rho^{(0)}_{jk}]+\ve\Big(\LI_1[\rho^{(0)}_{jk}]+\LI_0[\rho^{(1)}_{jk}]\Big)
\simeq-i(E_j-E_k)\rho^{(0)}_{jk}+\ve\Big(\eta_{jk}\,\rho^{(0)}_{jk}-i(E_j-E_k)\rho^{(1)}_{jk}\Big)\ ,
$$
whence
$$
-i[H\,,\,\rho^{(0)}_{jk}]=-i(E_j-E_k)\rho^{(0)}_{jk}\ ,\quad
-i[H\,,\,\rho^{(1)}_{jk}]+\LI_1[\rho^{(0)}_{jk}]=\eta_{jk}\,\rho^{(0)}_{jk}-i(E_j-E_k)\rho^{(1)}_{jk}\ .
$$
Setting $\rho^{(0)}_{jk}=\vert j\rangle\langle k\vert$, one gets
$$
\eta_{jk}=\langle j\vert\LI_1[\vert j\rangle\langle k\vert]\vert k\rangle=
\sum_\alpha\Big(\langle j\vert h_\alpha\vert j\rangle\langle k\vert h^\dag_\alpha\vert k\rangle
-\frac{1}{2}\Big(\langle j\vert h^\dag_\alpha h_\alpha\vert j\rangle+
\langle k\vert h^\dag_\alpha h_\alpha\vert k\rangle\Big)\Big)\ ,
$$
and, when either $i\neq j$ or $k\neq \ell$,
$$
\langle i\vert \rho^{(1)}_{jk}\vert \ell\rangle=\frac{\langle i\vert \LI_1[\vert j\rangle\langle k\vert]\vert \ell\rangle}
{i(E_i-E_\ell-E_j+E_k)}\ .
$$
Therefore,
\beqqn
\Re{\rm e}(\eta_{jk})&=&
\sum_\alpha\Big(\Re{\rm e}\Big(\langle j\vert h_\alpha\vert j\rangle\langle k\vert h^\dag_\alpha\vert k\rangle\Big)
-\frac{1}{2}\Big(\langle j\vert h^\dag_\alpha h_\alpha\vert j\rangle+
\langle k\vert h^\dag_\alpha h_\alpha\vert k\rangle\Big)
\Big)\\
&\leq&-\frac{1}{2}\sum_\alpha\left|\langle j\vert h_\alpha\vert j\rangle\,-\,
\langle k\vert h^\dag_\alpha\vert k\rangle
\right|^2\ .
\eeqqn
The above inequality is strict unless $\langle j\vert h^\dag_\alpha h_\alpha\vert j\rangle=|\langle j\vert h_\alpha\vert j\rangle|^2$ and same for $\langle k\vert h^\dag_\alpha h_\alpha\vert k\rangle$.
Then, $\Re{\rm e}(\eta_{jk})=0$ would imply that, for all $\alpha$,
$$
h_\alpha=\langle j\vert h_\alpha\vert j\rangle\,\Big(\vert j\rangle\langle j\vert\,+\,\vert k\rangle\langle k\vert \Big)\,+\,\sum_{i;\ell\neq j,k} \langle i\vert h_\alpha\vert \ell\rangle\, \vert i\rangle\langle \ell\vert\ ,
$$
whence, contrary to the assumptions, $h_\alpha$ reduces to a scalar multiple of $P=\vert j\rangle\langle j\vert\,+\,\vert k\rangle\langle k\vert$ on the subspace projected out by $P$.

Concerning the perturbation of the eigenvalue $0$, choose $\rho_0$ such that $\LI_0[\rho]=0$,
but $\wh{\DI}[\rho]\neq0$; then, to first order in $\ve$,
$$
(\LI_0+\ve\DI)[\rho+\ve\rho_1+o(\ve)]=\Big(\ve\alpha_1+o(\ve)\Big)\Big(\rho+\ve\rho_1+o(\ve)\Big)\Longrightarrow
\LI_0[\rho_1]+\DI[\rho]=\alpha_1\rho\ .
$$
By splitting $\DI[\rho]=\GI_0\circ\DI[\rho]+\FI_0\circ\DI[\rho]$, where $\FI_0={\rm id}-\GI_0$,
one gets the solutions $\rho_1=-\LI_0^{-1}\circ\FI_0\circ\DI[\rho]$ and $\wh{\DI}[\rho]=\alpha\rho$.
Since $\wh{\DI}$ generates a trace-preserving positive map on the $\gamma^{(0)}_t$-invariant states, the
eigenvalue $\alpha$ must be negative.
\medskip

\begin{exa}
\label{exx}
Consider $\DI$ as in Example~\ref{exa2}, where $h_\alpha=\vert\psi\rangle\langle\alpha\vert$.
Then, $[P\,,\,\vert\psi\rangle\langle\alpha\vert]=0$ for all $1\leq\alpha\leq d$ yields $P=1$.
\end{exa}

\subsection{Entanglement production}\label{entanglement production}

In this section we shall study whether an appropriate, purely dissipative Lindblad dynamics can create entanglement even when it is a weak perturbation of a non-entangling unitary dynamics. The fact that a Lindblad dynamics that does not include a unitary
part is able to create entanglement is shown in~\cite{BF1} in a very concrete example, and the fact that a unitary evolution can be added if the invariant state is an eigenstate of the unitary evolution is the result in~\cite{Kraus}. Here we concentrate on the assumption that this is exactly not the case; instead, we tackle the situation where the states invariant under the unitary time evolution are all separable. First, we observe that
\medskip

\begin{lem}
\label{noent}
Let the generator $\LI_0$ be given by a Hamiltonian of the form $H=H_1\otimes 1+1\otimes H_2$ where $H_1$ has eigenvalues $E_{1,k}$ and $H_2$ eigenvalues $E_{2,l}$ where $E_{1,k}\neq E_{2,l} \quad \forall l,k$.
Then, all $\gamma_t^{(0)}$-invariant states are separable. If the solutions of $\wh{\DI}[\rho]=0$, where $\wh{\DI}$ is as in\Ref{condexp2}, are not on the border of the set of separable states, then there exists $\ve_0$ such that for all $\ve\leq\ve_0$ the invariant state is unique and separable.
\end{lem}
\medskip

\noindent
\textbf{Proof:} Let $\ve_0$ be smaller than the radius in which the perturbation expansion of $\rho(\ve)$ such that $\LI_\ve[\rho(\ve)]=0$ converges. Because of Proposition~\ref{prop1}, if all solutions of $\wh{\DI}[\rho]=0$ are invertible, then there can exist only one; as a consequence (see the proof of Lemma~\ref{lem3}), there can be only one $\rho(\ve)$ within its convergence radius. Further, in a sufficiently small neighborhood of a state not on the border of the convex set of separable states, all states are separable.
\medskip

The following result will instead provide instances of a contrary behavior, more along the lines of Example~\ref{ex5}, showing the possibility of creating entanglement by weak dissipative perturbations.
\medskip

\begin{prp}
\label{prop2}
Consider the generator\Ref{pertLind} with a non-entangling Hamiltonian as in
Lemma \ref{noent} and a dissipative perturbation $\DI$ such that $\wh{\DI}[\rho]=0$ has only one solution. Then, there is a unique state in the kernel of $\LI_\ve$, given by the perturbation expansion
$\rho(\ve)=\sum_n\ve^n\rho_n$  where
\beq
\label{aita}
\rho_n=(-)^n\Big(
\Big({\rm id}-\wh{\DI}^{-1}\circ\GI_0\circ\DI\Big)\circ\LI_0^{-1}\circ\DI\Big)^n[\rho_0]\ .
\eeq
\end{prp}
\medskip

\noindent
\textbf{Proof:} Given a zero-th order approximation $\rho_0$ such that $\LI_0[\rho_0]=0$ and $\wh{\DI}[\rho_0]=0$,
we put ourselves in the most general situation where $\GI_0\circ\DI\circ\LI_0^{-1}\circ\DI[\rho_0]\neq0$.
We then add to $\rho_1=-\LI_0^{-1}\circ\DI[\rho_0]$ a matrix $\sigma_1$ such that $\LI_0[\sigma_1]=0$; the new matrix $\wt{\rho}_1=\rho_1+\sigma_1$ still solves $\LI_0[\wt{\rho}_1]=-\DI[\rho_0]$. Since we want to solve
$\LI_0[\rho_2]=-\DI[\wh{\rho}_1]$ by inverting $\LI_0$, we seek $\sigma_1$ such that (see\Ref{aitanos})
\beq
\label{aitanos1}
\GI_0\circ\DI[\rho_1]+\wh{\DI}[\sigma_1]=0\ .
\eeq
Since we assumed that $\wh{\DI}$ to have only one state in its kernel, it cannot vanish on $\GI_0\circ\DI[\rho_1]$;
the latter is a traceless matrix and both its normalized positive and negative parts would be states in the kernel of $\wh{\DI}$ (see the proof of Lemma~\ref{lem3}).
Therefore,
$$
\wt{\rho}_1=\Big({\rm id}-\wh{\DI}^{-1}\circ\GI_0\circ\DI\Big)[\rho_1]=
-\Big({\rm id}-\wh{\DI}^{-1}\circ\GI_0\circ\DI\Big)\circ\LI_0^{-1}\circ\DI[\rho_0]\ .
$$
Iterating this construction, one obtains the contributions to the perturbation expansion as in\Ref{aita}.
\medskip

\begin{rem}
\label{rem2}
In general, according to Lemma~\ref{lem2}, in order to solve equation\Ref{aitanos1} for a generic $\LI_1$ in the place
of $\DI$, one has to consider the map $\wh{\GI}_1$ that remains associated to it by the time-average\Ref{timeav}, and
check whether $\wh{\GI}_1\circ\GI_0\circ\DI[\rho_1]=0$.
\end{rem}
\medskip

\begin{exa}
\label{exa10}
The map $\wh{\DI}$ in Example~\ref{exa9} has only one invariant state; according to\Ref{exa5}
its inverse is given by $\wh{\DI}^{-1}[X]=-X$ on $X$ such that ${\rm Tr}(X)=0$.
Therefore, by means of\Ref{aitanos1} and\Ref{exa8.1}, equation\Ref{aita} with $n=1$  and
$\rho_0=\sum_{i=1}^d\rho_{ii}\,\vert i\rangle\langle i\vert$ yields
\beqq
\label{exa10.1}
\hskip -1cm
\rho_1&=&-\Big({\rm id}-\GI_0\circ\DI\Big)\circ\LI_0^{-1}\circ\DI[\rho_0]=
\sum_{i=1}^d|\psi(i)|^2\,\vert i\rangle\langle i\vert-i\ve\sum_{j\neq k}\frac{\psi(j)\psi^*(k)}{E_j-E_k}\,
\vert j\rangle\langle k\vert\\
\label{exa10.2}
\hskip -1cm
\rho(\ve)&=&(1+\ve)\,\sum_{i=1}^d|\psi(i)|^2\,\vert i\rangle\langle i\vert-i\ve\sum_{j\neq k}\frac{\psi(j)\psi^*(k)}{E_j-E_k}\,
\vert j\rangle\langle k\vert\,+\,o(\ve)\ .
\eeqq
Consider the bipartite setting of Lemma~\ref{noent} and set $1\leq\alpha,\beta\leq a$, $a^2=d$,
$$
\vert j\rangle=\vert\alpha\beta\rangle=\vert\alpha\rangle\otimes\vert\beta\rangle\qquad\hbox{where}\quad
H\vert\alpha\beta\rangle=E_{\alpha\beta}\vert\alpha\beta\rangle\ ,\ E_{\alpha\beta}=E_{1,\alpha}+E_{2,\beta}\ .
$$
By transposing the first party with respect to the orthonormal basis $\{\vert\alpha\rangle\}_{\alpha=1}^a$,
as in Example~\ref{ex5}, one obtains
$$
\rho^\Gamma(\ve)=(1+\ve)\,\sum_{\alpha,\beta=1}^a|\psi_{\alpha\beta}|^2\,
\vert \alpha\rangle\langle \alpha\vert\otimes\vert\beta\rangle\langle\beta\vert
-i\ve\sum_{(\alpha,\beta)\neq(\gamma,\delta)}\frac{\psi_{\alpha\beta}\psi_{\gamma\delta}^*}
{E_{\alpha\beta}-E_{\gamma\delta}}\,
\vert \gamma\rangle\langle \alpha\vert\otimes\vert\beta\rangle\langle\delta\vert\,+\,o(\ve)\ .
$$
Suppose $\psi_{\alpha_1\beta_1}=\psi_{\alpha_2\beta_2}=0$ for $\alpha_1\neq\alpha_2$ and
$\beta_1\neq\beta_2$; then, choosing an entangled state $\vert\phi\rangle$ supported only
in the subspace spanned by $\vert\alpha_1\beta_1\rangle$ and $\vert\alpha_2\beta_2\rangle$, one calculates
$$
\langle\phi\vert\rho^\Gamma(\ve)\vert\phi\rangle=
\ve\,\frac{\Im{\rm m}
\Big(\phi_{\alpha_1\beta_1}\psi_{\alpha_2\beta_1}\phi^*_{\alpha_2\beta_2}\psi^*_{\alpha_1\beta_2}
\Big)}{E_{1,\alpha_1}+E_{2,\beta_2}-\Big(E_{1,\alpha_2}+E_{2,\beta_1}\Big)}\,+\,o(\ve)\ .
$$
This expectation can always be made negative and thus, by applying the partial transposition criterion, $\rho(\ve)$ results entangled to order $\ve$. Notice that the assumptions on the coefficients ensure that the projection of
$\vert\psi\rangle$ onto the subspace spanned by $\vert\alpha_2\beta_1\rangle$ and $\vert\alpha_1\beta_2\rangle$
is entangled;
furthermore, Example~\ref{exx} ensures that $\rho(\ve)$ is an asymptotic state for the given time-evolution.
\end{exa}

\section{Summary}

A practical control of the invariant states of quantum dynamical semigroups generated by Lindblad type generators
is still partial, while this knowledge would very much be needed due to the fact that quantum information protocols might use suitably engineered open quantum time-evolutions to achieve entanglement asymptotically.

In this paper, we have considered an asymptotic approach to this problem and studied the fate of the separable invariant states of two non-interacting quantum systems when their Hamiltonian, unitary time-evolution is weakly perturbed by a Lindblad type dissipative contribution $\ve\,\LI_1$ to the generator $\LI_0$.

The investigation has been conducted by developing the first theoretical steps of a perturbative approach to the
stationary states $\rho(\ve)$ of the perturbed generator $\LI_\ve=\LI_0+\ve\,\LI_1$ and the preliminary results
have been applied to show how small suitable dissipative corrections to the unitary dynamics, not entangling by itself, may indeed lead to entangled asymptotic invariant states, $\LI_\ve[\rho(\ve)]=0$.

The main theoretical tool used to obtain these results has been the practical handling of the first steps of an iterative procedure that provides the contributions to the perturbative expansion of $\rho(\ve)$; the examples we have presented show that in some cases the iterative procedure can be performed. From a theoretical point of view, it remains to be understood whether the construction of the perturbative contributions to $\rho(\ve)$ can always be achieved by an appropriate choice of the zeroth order seed as in Section~\ref{iterative-sec} or whether the summation breaks at a finite order because of lack of analyticity.


\begin{thebibliography}{99}
\bibitem{AL}
R. Alicki, K. Lendi, {\em Quantum Dynamical Semigroups and
Applications}, Lect. Notes Phys. \textbf{717}, (Springer-Verlag,
Berlin, 2007).
\bibitem{BP}
H.--P. Breuer, F.Petruccione, {\it The Theory of Open Quantum Systems}
(Oxford University Press, Oxford, 2002).
\bibitem{Hororev}
R. Horodecki {\it et al.}, {\it Rev. Mod. Phys.} {\bf 81}, 865 (2009).
\bibitem{braun}
D. Braun, {\em Phys. Rev. Lett.} \textbf{89}, 277901 (2002).
\bibitem{beige}
A. Beige et al., {\em J. Mod. Opt.} \textbf{47}, 2583 (2000).
\bibitem{jakob1}
L. Jakobczyk, {\em J. Phys. A} \textbf{35}, 6383 (2002).
\bibitem{jakob2}
L. Jakobczyk, {\em J. Phys. B} \textbf{43}, 015502 (2010).
\bibitem{BFP}
F. Benatti, R. Floreanini, M. Piani, {\em Phys. Rev. Lett.} \textbf{91}, 070402 (2003).
\bibitem{BLN}
F. Benatti, A.M. Liguori, A. Nagy, {\em J. Math. Phys.} \textbf{49}, 042103 (2008).
\bibitem{Isar}
A. Isar, {\em Open Sys. Inf. Dynamics} \textbf{16}, 205 (2009).
\bibitem{BF1}
F. Benatti and R. Floreanini, {\em Int. J. Mod. Phys. B} \textbf{19}, 3063 (2005).
\bibitem{BF2}
F. Benatti, R. Floreanini, {\em J. Opt. B} \textbf{7}, S429 (2005).
\bibitem{B}
F. Benatti, {\em Lecture Note Series},
Institute for Mathematical Sciences, National University of Singapore \textbf{20}, 133 (2010).
\bibitem{BLPal}
F. Benatti, A.M. Liguori, G. Paluzzano, {\em J. Phys. A} \textbf{43}, 045304 (2010).
\bibitem{Kraus}
B. Kraus et al., {\em Phys. Rev. A} \textbf{78}, 042307 (2008).
\bibitem{Fri1}
A. Frigerio, {\em Lett. Math. Phys.} \textbf{2}, 79 (1977).
\bibitem{Fri2}
A. Frigerio, {\em Comm. Math. Phys.} \textbf{63}, 269 (1978).
\bibitem{spohn}
H. Spohn, {\em Lett. Math. Phys.} \textbf{2}, 33 (1977).
\bibitem{FagnRebol}
F. Fagnola, R. Rebolledo, {\em Lec. Notes in Math.} \textbf{1882}, 161 (2006).
\bibitem{Dietz}
K. Dietz, {\em J. Phys. A} \textbf{37}, 6143 (2004).
\bibitem{HBW}
B. Baumgartner, H. Narnhofer, W. Thirring, {\em J. Phys. A} \textbf{41}, 065201 (2008).
\bibitem{Umanita}
V. Umanit\`a, {\em Probab. Theory \& Related Fields} \textbf{134}, 603 (2006).
\bibitem{HB}
B. Baumgartner, H. Narnhofer, {\em J. Phys. A \textbf{41}}, 395303 (2008).
\end{thebibliography}
\end{document}